\documentclass[aps,prl,twocolumn,superscriptaddress,10pt]{revtex4-2}
\usepackage{graphicx}
\usepackage{bm}
\usepackage{xcolor}
\usepackage{ragged2e}
\usepackage{epstopdf}
\usepackage{booktabs}
\usepackage{multirow}
\usepackage[colorlinks=true,linkcolor=black,citecolor=blue,urlcolor=blue]{hyperref}
\usepackage{amsmath}
\usepackage{tabularx}
\usepackage{dcolumn}
\usepackage{cleveref}
\usepackage{easyReview}

\newcommand{\be}{\begin{equation}}
\newcommand{\ee}{\end{equation}}
\newcommand{\bea}{\begin{eqnarray}}
\newcommand{\eea}{\end{eqnarray}}

\newcommand{\bee}{\begin{eqnarray*}}
\newcommand{\eee}{\end{eqnarray*}}

\footnotetext{These authors contributed equally to this work.}
\crefname{equation}{Eq.}{Eqs.}
\crefname{figure}{Fig.}{Figs.}
\crefname{table}{Tab.}{Tabs.}
\begin{document}
\title{Quantitative measurement of fluid inertial effects in confined Brownian motion}
\author{Quentin Ferreira}
\affiliation{Univ. Bordeaux, CNRS, LOMA, UMR 5798, 33405 Talence, France}
\author{Pablo Palacios-Alonso$^a$}
\affiliation{Dpto. Física Teórica de la Materia Condensada, Universidad Autónoma de Madrid, 28049 Madrid, Spain}
\author{Harshit Joshi$^a$}
\affiliation{Univ. Bordeaux, CNRS, LOMA, UMR 5798, 33405 Talence, France}
\author{Rafael Delgado-Buscalioni}
\thanks{rafael.delgado@uam.es}
\affiliation{Dpto. Física Teórica de la Materia Condensada, Universidad Autónoma de Madrid, 28049 Madrid, Spain}
\affiliation{Condensed Matter Physics Center, IFIMAC, Spain}
\author{Yacine Amarouchene}
\thanks{yacine.amarouchene@u-bordeaux.fr}
\affiliation{Univ. Bordeaux, CNRS, LOMA, UMR 5798, 33405 Talence, France}
\author{Thomas Salez}
\thanks{thomas.salez@cnrs.fr}
\affiliation{Univ. Bordeaux, CNRS, LOMA, UMR 5798, 33405 Talence, France}
\affiliation{Mechanics Department, Ecole Polytechnique, Institut Polytechnique de Paris, 91128 Palaiseau, France}
\date{\today}
\begin{abstract}
The hydrodynamic response of Brownian particles in liquids is fundamentally altered by inertial forces arising from unsteady momentum transport in the surrounding fluid. These forces are of two distinct types: the added mass and the history effect. While both are well understood in bulk and weakly-confined geometries, under deterministic driving, their respective behaviours under strong confinement and thermal fluctuations remain scarcely addressed, unclear and often entangled together. 
	The goal of the present study is thus to fill this fundamental gap. The confinement-induced variations of the two distinct inertial contributions are separately investigated in the vicinity of a flat, rigid wall, using a combination of broadrange thermal colloidal-probe atomic-force-microscopy experiments, advanced numerical simulations and theoretical arguments. Specifically, the separation of the added-mass and history-force contributions is achieved through their different frequency-scaling signatures within the assumed form of the generalized dissipation coefficient.
	Our results pave the way towards a complete picture of Brownian motion at interfaces, in the lubrication regime, with direct relevance to nanofluidics and interfacial biophysics.
	\end{abstract}
\maketitle

Brownian motion is often described as memoryless: uncorrelated thermal fluctuations drive a suspended particle while viscous forces rapidly dissipate momentum into the surrounding fluid~\cite{Einstein1905,vonSmoluchowski1906,Langevin1908}. Yet, fluids do not relax momentum instantaneously. The unsteady motion of the particle generates vorticity that diffuses through the fluid over a finite time scale. When the latter becomes comparable to the particle's velocity-variation time scale, the resulting hydrodynamic response acquires memory~\cite{Boussinesq1885,Basset1888}. Therefore, fluid-inertial effects do matter -- even at vanishing Reynolds number.
Fluid inertial effects at low Reynolds number are separated in two distinct hydrodynamic mechanisms. First, the so-called \textit{added mass} originates from the instantaneous pressure field required to enforce incompressibility, and renormalizes the effective inertia of a particle moving in the surrounding fluid~\cite{Mo2015}. Second, the so-called \textit{history force} reflects the nonlocal diffusion of vorticity through the surrounding fluid, giving rise to colored thermal noise and long-lived velocity correlations~\cite{Ohbayashi1983, Berg-Sorensen2004, Berg-Sorensen2005, Franosch2011, Kheifets2014}. 
While the ballistic regime of a colloid in a liquid has been measured in the bulk recently for the first time~\cite{Huang2011}, in many microbiological and nanofluidic systems, Brownian motion occurs close to confining boundaries~\cite{Dasgupta2005,Block2007,Agayan2008,Hwang2019}, which have several consequences. For instance, conservative surface forces put aside, a rigid wall modifies viscous dissipation nearby~\cite{Brenner1961}, and thus colloidal mobility and displacement statistics~\cite{faucheuxconfined1994,bevanhindered2000,Lin2000,Dufresne2001,Banerjee2005,matse_test_2017,lavaud2021stochastic,altman2022,alexandre2023non}. Neighbouring boundaries also alter the propagation of vorticity through the fluid, thereby reshaping hydrodynamic memory~\cite{Gotoh1982,clarke2005drag,clarke2006three,basak2006hydrodynamic,Jeney2008,Simha2018}. Previous studies, using oscillating cantilevers~\cite{ClarkePRL} and optically-trapped colloids~\cite{Mo2015a}, showed that confinement can modify thermal spectra, suppress resonances in the positional power spectral density, decrease equilibrium velocity fluctuations~\cite{Mo2015}, and alter the long-time decay of velocity correlations~\cite{huang2015effect}.
Nevertheless, fluid-inertial effects in the strongly-confined lubrication regime remain poorly understood.
For instance, when a micrometric particle approaches a surface at nanometric separations, viscous dissipation becomes strongly localized within the thin intervening fluid layer. In contrast, because fluid inertia not only depends on local gap flow but also on global momentum transport through the surrounding fluid, matched asymptotic treatments accounting for outer-flow couplings~\cite{Chadwick2008}, potential-flow predictions for added-mass~\cite{Yang2010}, and local lubrication descriptions~\cite{Bigan2024} yield qualitatively different inertial responses. 
Lastly, although viscous and inertial effects have been probed experimentally with deterministic oscillations~\cite{Carpentier2015,Dupre2016}, or with thermal fluctuations~\cite{benmouna2002hydrodynamic,Zhang2023}, including in air and rarefied gases~\cite{Honig2010}, the separate contributions of the added mass and history force, as well as the deep-lubrication regime, have not yet been examined. We note that after submission of our manuscript, we discovered a recent and relevant theoretical study on a related problem~\cite{Sorkin2026}.

In the present study, we combine broad-range thermal colloidal-probe atomic-force-microscopy experiments, advanced numerical simulations and theoretical arguments, in order to address Brownian motion in the vicinity of a flat, rigid wall. Our central aim is to measure and understand how strong, nanometric confinement reshapes the fluid-inertial dynamics of colloidal particles. Importantly, we decouple the inertial hydrodynamic response into its distinct added-mass and history-force contributions, by harnessing their respective frequency scalings within the assumed form of the generalized dissipation coefficient. By analyzing Brownian fluctuations over broad temporal and spatial ranges, we characterize the evolution of the added mass and history force, from the bulk down to the deep lubrication regime. 
\begin{figure}[!t]
	\begin{center}
		\includegraphics[width=0.9\linewidth]{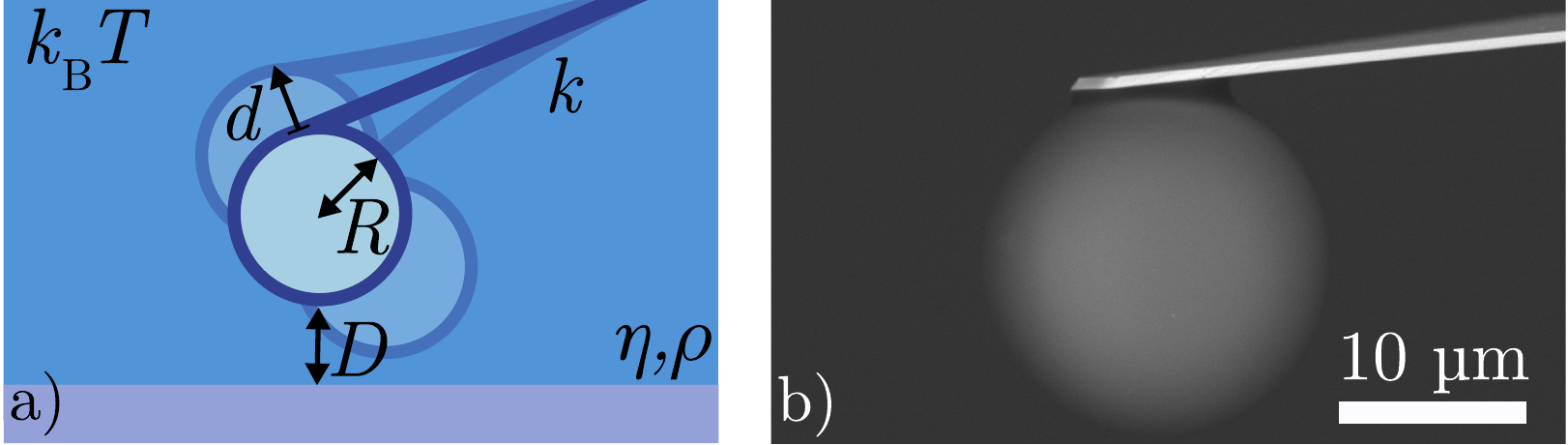}
		\caption{\textbf{Thermal atomic force microscopy colloidal-probe experiment near a rigid wall.} a) Schematic of the system. A micrometric sphere of radius $R$ is attached to an atomic force microscope cantilever of stiffness $k$, and placed in an incompressible Newtonian fluid of dynamic viscosity $\eta$ and density $\rho$ at temperature $T$. The sphere is further placed at an average distance $D$ from a planar rigid substrate, but undergoes Brownian motion due to thermal fluctuations. b) Corresponding scanning-electron-microscope image.}
		\label{fig:1}
	\end{center}
\end{figure}
\newline

We consider a borosilicate sphere of radius $R=8.85\pm0.15\,\mathrm{\mu m}$ in water, \textit{i.e.} an incompressible Newtonian fluid of dynamic viscosity $\eta$, density $\rho$, and thus kinematic viscosity $\nu=\eta/\rho$, at temperature $T$. The sphere is attached to an atomic force microscope (AFM) cantilever of stiffness $k$, positioned at an average distance $D$ from a planar solid borosilicate wall, and is let free to evolve within its fluid thermal bath (Fig.~\ref{fig:1}a). Note that a gold layer is present on the cantilever to increase its reflection coefficient. Such a thermal colloid-probe AFM setup enables robust continuous probing of inertial Brownian dynamics, through the stochastic deflection $d(t)$ of the cantilever along time $t$, from the bulk regime ($D\gg R$) down to the deep-lubrication regime ($D\ll R$). The displacement of the probe remains small ($d\ll R$), placing the system in a regime where convective inertial effects are negligible, with the Reynolds number $\mathrm{Re}\sim O(10^{-4})$. In contrast, unsteady fluid inertial effects matter, as evaluated by the $\sim O(1)$ Womersley number, $\mathrm{Wo}\equiv \sqrt{f/f_\nu}$, where $f_\nu^{-1}\equiv \pi R^2/\nu$ is the vorticity diffusion time scale, and $f$ is the frequency variable associated with the colloidal motion. In addition to experiments, numerical simulations are performed with a recently-developed spectral fluid solver for a bounded domain, coupled with the vibrational dynamics of viscoelastic immersed structures modelled as discrete beads~\cite{uammd2025,Pelaez2025,Palacios-Alonso2025,Bonet2026}. Details on the experimental and numerical methods are provided in the Supplementary Information (SI).

The central observable in both the experiments and simulations is the complex friction coefficient $\gamma$, which encompasses the whole, frequency-dependent and space-dependent, hydrodynamic action of the surrounding fluid onto the sphere-cantilever probe. Inspired by the expression of the classical hydrodynamic force acting on a rigid sphere in an unbounded fluid~\cite{Bedeaux1974,Landau2013}, we express the complex friction coefficient as:
\begin{equation}
	\label{eq:gamma_bulk}
	\gamma=\gamma_\mathrm{St}\left[c_0 + c_\mathrm{B}(1-i)\sqrt{\frac{f}{f_\nu}} - ic_{\textrm{m}}\left(\frac{f}{f_\nu}\right) \right],
\end{equation}
where $\gamma_\mathrm{St}=6\pi\eta R$ is the Stokes friction coefficient, and the three dimensionless coefficients, $c_0$, $c_\mathrm{B}$, and $c_{\textrm{m}}$, quantify respectively the steady-viscous, history-force, and added-mass contributions of the sphere-cantilever probe as a whole. 
The main hypothesis of the current study is that the presence of a neighbouring flat rigid wall preserves such a decomposition, while rendering the three coefficients distance-dependent, \textit{i.e.} $c_0(D/R)$, $c_\mathrm{B}(D/R)$, and $c_{\textrm{m}}(D/R)$. 

In the numerical simulations (see details in SI), $\gamma(f, D/R)$ is inferred from the oscillatory velocity response of the object to a prescribed external force, while, in the experiments, $\gamma(f, D/R)$ is extracted from the Brownian fluctuations of the probe through the positional power spectral density (PSD). For a thermally-driven harmonic oscillator, the one-sided PSD is given by~\cite{Berg-Sorensen2004}:
\begin{equation}
	\label{eq:PSD_generic}
	\mathrm{PSD}(f)=\frac{4k_\mathrm{B}T\mathrm{Re}(\gamma)}{\left| k-2\pi i f \gamma- (2\pi f)^2 m \right|^2}~,
\end{equation} 
where $k_{\mathrm{B}}$ is the Boltzmann constant, and $m$ is the dry mass. The latter corresponds to the dry mass of the sphere-cantilever probe in the experiments. We stress that $k$ does not correspond the steady probe stiffness, but to the effective modal stiffness that depends in particular on viscous dissipation and mode coupling~\cite{Sader1998,Yamamoto2007,Bellon2008}. Details on the determination of the effective stiffness $k$ are provided in the SI.
\begin{figure}[!t]
	\begin{center}
		\includegraphics[width=0.9\linewidth]{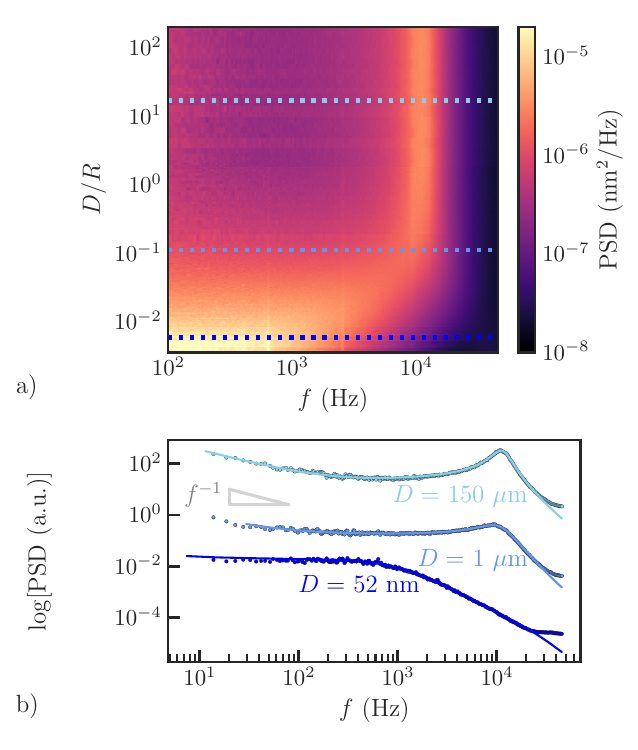}
		\caption{\textbf{Influence of the wall on the thermal spectrum of the colloidal probe.} a) Experimental positional power spectral density (PSD) of the sphere-cantilever probe in water as a function of the frequency $f$ and normalized distance $D/R$ to the wall. b) Experimental PSDs (points), in arbitrary units (a. u.) and vertically shifted for clarity, as functions of the frequencies $f$, for three specific distances $D$, as indicated, and corresponding to the dotted lines in panel a). In addition, the respective fits to \cref{eq:PSD_generic} (solid lines) are shown. The fitting is performed over a frequency range between $f_{\min}\in\left[1,80\right]\,\mathrm{Hz}$ and  $f_{\max}=46\,\mathrm{kHz}$, covering the first mode of the oscillator, with minimal interferences from the higher-order modes and experimental noise level. The slope triangle indicates the trend of the $1/f$ noise visible at low frequencies and powers.}
		\label{fig:2}
	\end{center}
\end{figure}

The experimental PSDs are plotted in \cref{fig:2}a), for a wide range of separations, from $D\sim2\,\mathrm{mm}$ down to $\sim35\,\mathrm{nm}$, which span the full crossover from the bulk regime to the lubrication regime. In order to filter the raw periodograms, Welch's method~\cite{Welch1967} with a window of $2.5\times10^4$ points and 50\% overlap was used. Also, to balance the fitting weights of low and high frequencies, the filtered PSDs have been block-averaged following a logarithmic sequence. The spectra in \cref{fig:2} reveal a crossover around $D/R\sim0.1$ from an underdamped regime to an overdamped one. Far from the wall, the resonance is readily identified near $11\,\mathrm{kHz}$. As the probe approaches the substrate, the PSD broadens as a consequence of the increased dissipation, and the resonance progressively disappears.

\begin{figure}[!t]
	\begin{center}
		\includegraphics[width=0.9\linewidth]{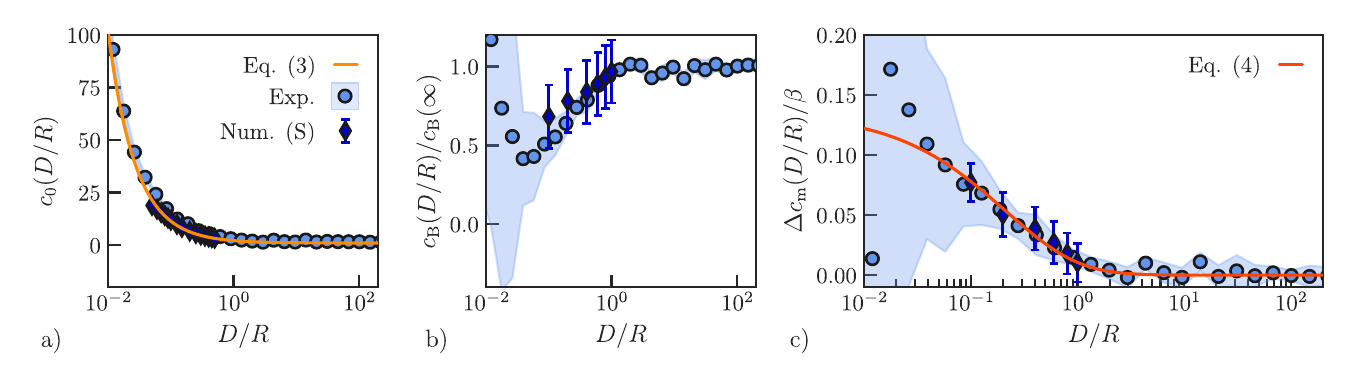}
		\caption{\textbf{Influence of the wall on viscous damping and fluid inertial effects.} a) Experimental steady-viscous contribution $c_0$ (discs) of the sphere-cantilever probe, as a function of normalized distance $D/R$. Also shown are the corresponding numerical data (diamonds) for a pure sphere. The solid line represents Brenner's prediction  for a pure sphere (see \cref{eq:PadeBrenner})~\cite{Brenner1961,Bevan2000}. b) Experimental history-force contribution $c_\mathrm{B}$ (discs) of the sphere-cantilever probe, normalized by its bulk value $c_\mathrm{B}(\infty)$, as a function of normalized distance $D/R$. Also shown are the corresponding numerical data (diamonds) for a pure sphere. c) Experimental increase $c_\mathrm{m}-c_\mathrm{m}(\infty)$ in the added-mass contribution (discs) of the sphere-cantilever probe, with respect to the bulk value, as a function of normalized distance $D/R$. The data is normalized by the geometry-accommodation factor $\beta=18$ (see text). Also shown are the corresponding numerical data (diamonds) for a pure sphere with $\beta=1.6$. The solid line represents Yang's prediction  for a pure sphere (see \cref{eq:YangMass}). The experimental data in the three panels was obtained from fitting the PSDs to \cref{eq:gamma_bulk,eq:PSD_generic} (see \cref{fig:2}b)). 
		To improve readability, the data has been log-blocked using six blocks per decade in $D/R$. The associated error bands represent the maxima between the $2\sigma$-deviations of the values within a block and the averages of the $2\sigma$ fit uncertainties.}
		\label{fig:3}
	\end{center}
\end{figure}

Using \cref{eq:PSD_generic} together with \cref{eq:gamma_bulk}, one can properly fit the experimental PSDs, as shown in \cref{fig:2}. Three independent fit parameters are identified: $c_0$, $c_\mathrm{B}$ and $c_{\textrm{m}}$. Note that, for the latter, we rather use the effective mass $m+\gamma_\mathrm{St}c_{\textrm{m}}/(2\pi f_\nu)$ as a fit parameter in practice. Additionally, a fourth free parameter $c$, accounting for the systematic $1/f$ noise, is introduced. As expected, $c$ is found to be essentially independent of $D$. The main fit parameters, $c_0(D/R)$, $c_\mathrm{B}(D/R)$, and $c_{\textrm{m}}(D/R)$, for the experimental sphere-cantilever probe are shown in \cref{fig:3}, together with the corresponding results from numerical simulations for a pure sphere (see SI for the numerical study of a sphere-cantilever geometry), denoted by $c_0^\mathrm{(S)}(D/R)$, $c_\mathrm{B}^\mathrm{(S)}(D/R)$, and $c_{\textrm{m}}^\mathrm{(S)}(D/R)$. Below, we discuss these three quantities.

Let us start first with the steady-viscous contribution. The viscous drag coefficient $\gamma_\mathrm{St}c_0^\mathrm{(S)}$ for a pure sphere moving perpendicularly to a rigid, no-slip and flat wall, was theoretically derived by Brenner~\cite{Brenner1961}. A compact Padé approximation~\cite{Bevan2000} of this result is given by:
\begin{equation}\label{eq:PadeBrenner}
  c_0^\mathrm{(S)}(D/R)=\frac{6D^2/R^2+9D/R+2}{6D^2/R^2+2D/R}~.
\end{equation} 
As seen in \cref{fig:3}a), \cref{eq:PadeBrenner} matches the near-wall experimental and numerical data, with no free parameter. This is expected since, near the wall, the contribution of the sphere dominates through the increased dissipation localized within the sphere-wall lubrication gap. In contrast, far away from the wall, the experimental steady viscous friction results from an intricate combination of the cantilever and sphere geometries, with a bulk drag coefficient value higher than that for a pure sphere. This is a classical observation in bulk colloidal-probe AFM experiments. Besides, hydrodynamic corrections taking into account surface roughness are discussed in the SI. 

Let us now turn to the fluid inertial effects. Prior to that, we stress that the presence of a nearby boundary is expected to affect inertial effects, as for the steady-viscous contribution above. However, the lubrication gap is not necessarily the only dominant region anymore for inertial effects near a wall, and the whole sphere-cantilever geometry matters. In fact, one of the main points of our study is precisely that we do not discard the influence of the cantilever in our analysis, as its hydrodynamic contribution to the power spectra is an indisputable fact in any AFM experiment. Trying to decouple the sphere and cantilever's contributions would in contrast lead to coarse, uncontrolled assumptions on the nature of their hydrodynamic coupling. 

We first consider the history-force contribution. Figure~\ref{fig:3}b) shows $c_\mathrm{B}(D/R)$ normalized by its bulk value $c_\mathrm{B}(\infty)$. Interestingly, the otherwise-important geometrical differences are absorbed by such a normalization choice, so that the experimental and numerical data agree with each other. However, unlike the steady-viscous contribution above, a closed-form analytical expression does not exist in our parameter range, even for a pure sphere. Overall, within our error bars, a decrease in the history-force contribution is observed with decreasing $D/R$.
This observation is consistent with previous studies that have reported a wall-induced suppression of hydrodynamic memory~\cite{Gotoh1982,Jeney2008,Mo2015a,Huang2015}.
Besides, we stress that the uncertainty in the fit parameter $c_\mathrm{B}(D/R)$ increases significantly for $D/R\lesssim0.1$, which may be due to the steady-viscous contribution dominating the hydrodynamic response, and hence the overall shape of the PSD (see SI for details on the increasing errors near the wall). 

\onecolumngrid
~    
\twocolumngrid
Finally, let us examine the added-mass contribution. For a pure sphere accelerating perpendicularly to a flat rigid wall, $c_{\textrm{m}}^\mathrm{(S)}(D/R)$ can be obtained using potential-flow (\textit{i.e.} inviscid) theory, and an expression for all $D/R$ has been derived by Yang~\cite{Yang2010}, based on previous works~\cite{Lamb1945, MilneThomsonL.N.1962}. It reads:
\begin{multline}\label{eq:YangMass}
  c_{\textrm{m}}^\mathrm{(S)}(D/R)= \frac{2}{9} \left[1 + 3 \sum_{n=1}^\infty\left( \zeta^{-n/2} \sum_{k=0}^n \zeta^k \right)^{-3} \right],\\ \text{with}\quad \zeta \equiv \frac{1+D/R-\sqrt{(1+D/R)^2-1}}{1+D/R+\sqrt{(1+D/R)^2-1}}.
\end{multline}
Far from the wall, $c_{\textrm{m}}^\mathrm{(S)}$ reaches the expected bulk value $c_{\textrm{m}}^\mathrm{(S)}(\infty)=2/9$, while it equals $c_{\textrm{m}}^\mathrm{(S)}(0)\approx 0.3569$ at the wall. Interestingly, the latter is larger than the bulk value, hence indicating an enhancement of added-mass effects near a flat rigid wall. Since the experimental data for $c_{\textrm{m}}(D/R)$ corresponds to the added mass of the full sphere-cantilever probe, it is not directly comparable to $c_{\textrm{m}}^\mathrm{(S)}$ from \cref{eq:YangMass}. Nevertheless, we propose an Ansatz, where the increase in added mass with respect to the bulk value is described by a modified version of \cref{eq:YangMass} using a constant geometry-accommodation factor $\beta$, as follows: $c_{\textrm{m}}(D/R)-c_{\textrm{m}}(\infty) \approx \beta[c_{\textrm{m}}^\mathrm{(S)}(D/R)-c_{\textrm{m}}^\mathrm{(S)}(\infty)]$. We stress that $\beta$ can be as large as required for the considered specific geometry, and as such should not be interpreted as a scaling prefactor of order unity. In \cref{fig:3}c), the difference between $c_{\textrm{m}}(D/R)$ and its bulk value $c_{\textrm{m}}(\infty)$ is plotted. As observed, the Ansatz works with $\beta=18$. Once again, we see that a proper rescaling allows to absorb the non-spherical geometrical corrections, which are inherent to fluid inertial effects in colloidal-probe AFM experiments. As expected, the $\beta$ value is much larger than 1, indicating enhanced added-mass effects with respect to the pure-sphere case due to the presence of the cantilever. This is qualitatively confirmed by a similar analysis performed on numerical-simulation results for a cantilever-sphere system (see SI). For comparison, the numerical data for a pure sphere is also shown here. We stress that $\beta=1.6$ is found in the latter case, which slightly differs from $1$, possibly resulting from confinement-induced departures from Eq.~(\ref{eq:gamma_bulk}) for the history-force contribution~\cite{Felderhof2005,Felderhof2012,Simha2018}. Finally, we note that an increase of error bars in $c_{\textrm{m}}$, similar to the one on $c_{\textrm{B}}$, is observed at the smallest accessed $D$ values (see SI for details).

In summary, using a combination of experiments, numerical simulations and theoretical arguments, we have estimated for the first time the changes in both the added mass and history force acting on a Brownian particle, in close proximity to a rigid flat wall. Under an assumed form of the generalized dissipation coefficient, the positional PSD serves as a suitable observable, enabling the measurement of the full hydrodynamic response, from the bulk down to the lubrication regime. In the deep lubrication regime, the steady viscous drag component dominates the response within the frequency domain corresponding to the first mode of the oscillator. This may explain the loss of observability of the history force and added mass at these extreme confinement scales. 
Moreover, the simple bulk-like $\sim\sqrt{f}$ scaling hypothesis for the history-force contribution within the assumed form of the generalized friction coefficient requires further investigation. Indeed, previous studies~\cite{Felderhof2005,Felderhof2012,Jeney2008,Mo2015a,Huang2015,Simha2018,Fouxon2018a} suggest a more complex frequency dependence in the presence of a nearby wall -- despite the lack of any exact, analytical expression describing such a dependence. Such a modification of the generalized friction coefficient may alter our results at the smallest distances and call for further explorations. Lastly, the present study sets the ground for future investigations of Brownian objects interacting with soft boundaries, where elastohydrodynamic and viscoelastic effects are expected to play a central role~\cite{Palacios-Alonso2025, Bonet2026}. Previous studies already suggest that inertial effects could be essential to understand these complex coupled systems, whether for the improvement of microrheological techniques~\cite{Indei2012,Makris2021}, or contactless AFM~\cite{Zhang2025}, or for explaining the thermal diffusion behaviour in biologically-relevant systems~\cite{Marbach2018,Fares2024,Ye2025}. The proper separation of history-force and added-mass contributions seems essential for segmentation of the different effects that may emerge in such situations.

\subsection*{Data availability}
The data and codes are available at  
\url{https://github.com/EMetBrown-Lab/fluid-inertia-AFM} for experiments, and \url{https://uammd.readthedocs.io/en/latest/} for numerical simulations.

\section{acknowledgments}
The authors thank Vincent Bertin, Pascal Damman, Carlos Drummond, Nicolas Fares and Quentin Thomas for interesting discussions. They acknowledge financial support from the European Union through the European Research Council under EMetBrown (ERC-CoG-101039103) grant. They also acknowledge financial support from the University of Bordeaux Prematuration under BioFiX grant, and from Agence Nationale de la Recherche under EMetBrown (ANR-21-ERCC-0010-01), Softer (ANR21-CE06-0029), and Fricolas (ANR-21-CE06-0039) grants, as well as from the Interdisciplinary and Exploratory Research program under MISTIC grant at the University of Bordeaux. In addition, they acknowledge funding from the Spanish Agencia Estatal de Investigacion from project PID2024-158994OB-C41. Besides, they acknowledge the support from the R\'eseaux de Recherche Impulsion Frontiers of Life which received financial support from the French government in the framework of the University of Bordeaux's France 2030 program. Finally, they thank the Soft Matter Collaborative Research Unit, Frontier Research Center for Advanced Material and Life Science, Faculty of Advanced Life Science at Hokkaido University, Sapporo, Japan, and the CNRS International Research Network between France and India on Hydrodynamics at small scales: From soft matter to bioengineering.
\bibliography{Ferreira2026}
\end{document}